\documentclass[pra,twocolumn,superscriptaddress,longbibliography]{revtex4-2}
\usepackage[colorlinks=true, citecolor=blue, urlcolor=blue, linkcolor=blue ]{hyperref}
\usepackage{empheq,amssymb,orcidlink}

\begin{document}
\title{Extreme volume monogamy via bound-state engineering}
\author{Wen-Jie Zhang}
\affiliation{Key Laboratory of Quantum Theory and Applications of MoE, Lanzhou Center for Theoretical Physics, Gansu Provincial Research Center for Basic Disciplines of Quantum Physics, Key Laboratory of Theoretical Physics of Gansu Province, Lanzhou University, Lanzhou 730000, China}
\author{Jun-Hong An\orcidlink{0000-0002-3475-0729}}\email{anjhong@lzu.edu.cn}
\affiliation{Key Laboratory of Quantum Theory and Applications of MoE, Lanzhou Center for Theoretical Physics, Gansu Provincial Research Center for Basic Disciplines of Quantum Physics, Key Laboratory of Theoretical Physics of Gansu Province, Lanzhou University, Lanzhou 730000, China}
	
\begin{abstract}
Quantum steering ellipsoid (QSE) provides a faithful representation of a two-qubit state. When extended to tripartite systems, the steerability from a trusted party to different receivers is subject to volume monogamy relations, which only constrain the total steerability but cannot individually eliminate the steerability of an untrusted third party, leaving a potential channel for information leakage via steering. Here, we show that this residual steerability can be completely suppressed by selectively engineering bound states in local qubit-environment subsystems, without compromising the steerability between trusted parties. Specifically, when bound states are formed in the subsystems formed by the trusted parties and their environments but absent in the untrusted one, the untrusted party’s QSE volume decays to zero, while the trusted party’s QSE volume remains finite. Our results establish selective bound-state engineering as a mechanism for extreme volume monogamy, with potential applications in secure quantum communication with an untrusted third party.
\end{abstract}
\maketitle
\section{Introduction}

Quantum steering, a quantum-correlation phenomenon first identified by Schr\"odinger \cite{Schrödinger_1935,Schrödinger_1936}, allows one party to remotely alter the state of another spatially separated party via local measurements and has been formalized through diverse theoretical frameworks \cite{PhysRevLett.98.140402,saunders_experimental_2010,PhysRevLett.130.221801,RevModPhys.92.015001,PhysRevLett.115.230402,PhysRevA.40.913,RevModPhys.89.015002,PhysRevLett.48.291,PhysRevLett.115.210401,PhysRevLett.122.240401,PhysRevX.5.041008,PhysRevLett.70.1895}. For a two-qubit state, the steerability of one party by the other is fully characterized by the quantum steering ellipsoid (QSE), defined as the complete set of Bloch vectors realizable for the steered qubit via all possible positive-operator-valued measurements (POVMs) on the steering party's qubit \cite{PhysRevLett.113.020402}. Analogous to the Bloch sphere’s role in describing single qubits \cite{Nielsen_Chuang_2010}, the QSE translates the abstract structure of a two-qubit state into a tangible geometric form, rendering critical properties such as the nested tetrahedron condition for separability directly observable \cite{PhysRevLett.113.020402, Milne_2014}. The QSE thus captures the intricate geometry of two-qubit states and serves as a visual tool for diagnosing multiple forms of quantum correlation, including quantum discord \cite{PhysRevLett.113.020402,PhysRevA.91.022301,PhysRevA.85.064104,Shi_2011}, entanglement \cite{PhysRevLett.113.020402,PhysRevA.94.012114,Milne_2014}, EPR steering \cite{PhysRevA.94.012114,Jevtic_2015,PhysRevA.95.012320,Song_2023,PhysRevA.97.022338,PhysRevA.95.042117}, and Bell nonlocality \cite{PhysRevA.90.024302}. 

It is well known that quantum correlations cannot be freely shared among multiple parties, a feature captured by various monogamy relations for discord \cite{PhysRevA.86.062106,PhysRevA.84.054301,PhysRevB.84.245133}, entanglement \cite{PhysRevA.61.052306,PhysRevLett.96.220503,PhysRevA.69.022309,PhysRevA.89.034303,PhysRevA.80.044301,PhysRevA.77.032329,PhysRevA.99.042305,PhysRevA.79.012329,PhysRevLett.114.140402}, EPR steering \cite{PhysRevA.98.052325,PhysRevA.105.012202,PhysRevA.95.010101,PhysRevA.101.053834,PhysRevA.110.012418,PhysRevA.88.062108,PhysRevLett.128.120402,PhysRevLett.118.230501,PhysRevA.105.042435,PhysRevResearch.2.032046,PhysRevA.89.022332}, and Bell nonlocality \cite{PhysRevLett.87.117901,PhysRevLett.106.180402,PhysRevLett.112.100401,PhysRevLett.97.170409,PhysRevA.95.030104}. These monogamy relations are not only of fundamental interest in quantum physics but also play important roles in other areas, such as no-signaling theories \cite{PhysRevLett.109.050503,PhysRevA.90.052323}, condensed-matter physics \cite{ma_quantum_2011,PhysRevB.87.085130,PhysRevLett.116.130501}, and black-hole physics \cite{lloyd_unitarity_2014}. Beyond bipartite diagnostics, the QSE framework provides a tool for characterizing monogamy in multipartite systems. Milne \textit{et al.} derived a volume monogamy relation for pure three-qubit states \cite{Milne_2014}, later extended to mixed states \cite{PhysRevA.94.042105} and experimentally verified \cite{PhysRevLett.122.070402}. In realistic quantum networks, a tripartite entangled state may already be distributed among Alice, Bob, and Charlie, with Alice and Bob trusted and Charlie untrusted. Taking Alice as an example, the steerability of Bob and Charlie by Alice is quantified by their QSE volumes, which reflect how much information each party can access. However, conventional volume monogamy only bounds the total steerability and cannot fully eliminate Charlie's residual steerability \cite{Milne_2014,PhysRevA.94.042105}. A nonzero QSE volume of Charlie indicates steerability-mediated information accessibility, i.e., such steerable correlations could in principle be exploited by an untrusted party. If this volume vanishes, there is no steerable correlation left for the untrusted party to exploit. On the other hand, environmental decoherence degrades the correlations between trusted parties and shrinks their QSEs \cite{breuer_theory_2007,PhysRevA.102.020402}, destroying the basis for secure communication. This motivates us to explore extreme monogamy in quantum systems, i.e., eliminating the untrusted party’s steerability-mediated information accessibility while preserving the steerability essential for quantum communication between trusted parties by decoherence effect.

In this work, we present an approach to achieve extreme volume monogamy in a tripartite open quantum system. We consider three qubits, each coupled to its own local environment, with Alice and Bob trusted and Charlie untrusted. Our analysis shows that the geometry of each party’s QSE is determined by non-Markovian effects and the presence of bound states of the qubit and its environment. By controlling the formation of bound states, we can selectively preserve or eliminate the steerability of each party. When bound states are formed on the trusted parties but absent on the untrusted one, the untrusted party’s QSE volume decays to zero while the trusted party’s QSE volume remains finite. We systematically examine four scenarios: three-sided, two-sided, one-sided, and no bound states, which yield a rich variety of results. This work provides a physical basis for secure communication with an untrusted third party.

The remainder of this paper is organized as follows. In Sec. \ref{1}, we introduce the QSE framework and review the volume monogamy relations. In Sec. \ref{2}, we analyze how volume monogamy behaves in noisy environments, showing its connection to bound-state formation in each qubit-environment subsystem. In Sec. \ref{3},we present numerical results for different bound-state configurations, demonstrating the realization of extreme volume monogamy. Finally, we summarize our conclusions in Sec.~\ref{4}.

\section{QSE AND Volume monogamy}\label{1}

Consider a bipartite qubit state $\rho_{XY}$ shared by two parties, $X$ and $Y$. In the Bloch representation, it can be expressed as $ \rho_{XY} = \frac{1}{4}( \mathbb{I}_X \otimes \mathbb{I}_Y + \mathbf{x} \cdot \boldsymbol{\sigma} \otimes \mathbb{I}_Y + \mathbb{I}_X \otimes \mathbf{y} \cdot \boldsymbol{\sigma} + \sum_{j,k=1}^{3} T_{jk} \, \sigma_j \otimes \sigma_k )$, where $\mathbb{I}_{X,Y}$ are identity operators, $\boldsymbol{\sigma} = (\sigma_1, \sigma_2, \sigma_3)$ is the Pauli vector. $\mathbf{x}$ and $\mathbf{y}$ are the Bloch vectors of the reduced states of $X$ and $Y$ with components $x_j = \mathrm{Tr}[\rho_{XY}(\sigma_j \otimes \mathbb{I}_Y)]$ and $y_k = \mathrm{Tr}[\rho_{XY}(\mathbb{I}_X \otimes \sigma_k)]$, respectively. $T_{jk} = \mathrm{Tr}[\rho_{XY} (\sigma_j \otimes \sigma_k)]$ are the components of the spin correlation matrix \cite{PhysRevA.54.1838}. When party $X$ performs a local measurement described by a POVM element $E = e_0(\mathbb{I}_X + \mathbf{e} \cdot \boldsymbol{\sigma})$ with $0 \le e_0 \le 1$ and $|\mathbf{e}| \le 1$, party $Y$'s qubit is steered to the state $\rho_Y^E = 1/2[ \mathbb{I}_Y + \frac{\mathbf{y} + T^{\mathsf{T}} \mathbf{e}}{1 + \mathbf{x} \cdot \mathbf{e}} \cdot \boldsymbol{\sigma} ]$, with probability $p^E = e_0(1 + \mathbf{x} \cdot \mathbf{e})$. Then, considering all possible local measurements by $X$, it follows that the corresponding set of $Y$'s steered states is represented by the set of Bloch vectors \cite{PhysRevA.94.042105,PhysRevLett.113.020402}
\begin{equation}
    \mathcal{E}_{Y|X} = \left\{ \frac{\mathbf{y} + T^{\mathsf{T}} \mathbf{e}}{1 + \mathbf{x} \cdot \mathbf{e}} : |\mathbf{e}| \le 1 \right\}.
\end{equation}
This set forms an ellipsoid in the Bloch ball, known as the quantum steering ellipsoid $\mathcal{E}_{Y|X}$ \cite{PhysRevLett.113.020402}. The subscript $Y|X$ indicates the steering ellipsoid for $Y$ generated by $X$'s local measurements. The ellipsoid $\mathcal{E}_{Y|X}$ is fully determined by its center
\begin{equation}
    \mathbf{c}_{Y|X} = \frac{\mathbf{y} - T^{\mathsf{T}} \mathbf{x}}{1 - x^2}, \quad x = |\mathbf{x}|,
\end{equation}
and the orientation matrix
\begin{equation}
    Q_{Y|X} = \frac{1}{1 - x^2} (T - \mathbf{x} \mathbf{y}^{\mathsf{T}})^{\mathsf{T}} \left( \mathbb{I} + \frac{\mathbf{x} \mathbf{x}^{\mathsf{T}}}{1 - x^2} \right) (T - \mathbf{x} \mathbf{y}^{\mathsf{T}}).
\end{equation}
The eigenvalues of $Q_{Y|X}$ give the squared lengths of the ellipsoid's semiaxes, denoted $(r_{Y|X}^{x,y,z})^2$, and determine their orientations. The quantum steering ellipsoid $\mathcal{E}_{Y|X}$, along with the local Bloch vectors $\mathbf{x}$ and $\mathbf{y}$, offers a faithful geometric representation of the two-qubit state $\rho_{XY}$ \cite{PhysRevLett.113.020402}. A convenient measure of the ellipsoid's size is its normalized volume \cite{PhysRevLett.113.020402},
\begin{equation}
    v_{Y|X} = \frac{|\det(T - \mathbf{x} \mathbf{y}^{\mathsf{T}})|}{(1 - x^2)^2},
\end{equation}
where the normalization is taken with respect to the total volume of the Bloch sphere, $4\pi/3$. Now consider a tripartite scenario where Alice, Bob, and Charlie share a three-qubit state. By performing measurements on Alice's qubit, she can steer the states of Bob and Charlie. The corresponding sets of steered states form the steering ellipsoids $\mathcal{E}_{B|A}$ and $\mathcal{E}_{C|A}$, whose sizes are quantified by the normalized volumes $v_{B|A}$ and $v_{C|A}$, respectively. For a pure three-qubit state $|\psi_{ABC}\rangle$, when Alice performs measurements on her qubit, the normalized volumes of the QSE satisfy the monogamy relation $\sqrt{v_{B|A}} + \sqrt{v_{C|A}} \le 1$ \cite{PhysRevLett.122.070402}. Similarly, when Bob is the measured party, the analogous relation $\sqrt{v_{A|B}} + \sqrt{v_{C|B}} \le 1$ also holds \cite{Milne_2014}. For mixed states, however, these inequalities are not always valid. Instead, a weaker monogamy relation $(v_{B|A})^{2/3} + (v_{C|A})^{2/3} \le 1$ that holds for all three-qubit states has been derived in Ref. \cite{PhysRevA.94.042105}. 

\section{VOLUME MONOGAMY in noisy environments}\label{2}

We now consider that, among the three spatially separated parties, Alice and Bob are trusted parties, while Charlie is untrusted. Their qubits are independently coupled to their own environments. Their qubits are independently coupled to their own zero-temperature bosonic environments, each modeled as a bath of harmonic oscillators. The total Hamiltonian of the system is $\hat{H}_{\mathrm{total}} = \hat{H}_A + \hat{H}_B + \hat{H}_C$, where for $j = A, B, C$,
\begin{equation}
\hat{H}_j = \omega_{0} \hat{\sigma}_j^\dagger \hat{\sigma}_j + \sum_{k} [ \omega_{k} \hat{a}_{jk}^{\dagger} \hat{a}_{jk} + ( g_{jk} \hat{\sigma}_j^\dagger \hat{a}_{jk} + \mathrm{h.c.} ) ].\label{m3}
\end{equation}
Here $\hat{\sigma}_j = |g_j\rangle \langle e_j|$ is the lowering operator of the $j$th qubit, with transition frequency $\omega_0$. The operators $\hat{a}_{jk}^{\dagger}$ and $\hat{a}_{jk}$ are the creation and annihilation operators of the $k$th mode of the environment coupled to the $j$th qubit, with frequency $\omega_k$. In the continuum limit, the spectral density of each environment is given by $J_j(\omega) = \sum_k |g_{jk}|^2 \delta(\omega - \omega_{jk})$. We assume that the environments coupled to different qubits are independent, each described by an Ohmic-family spectral density $J_j(\omega) = \eta_j \omega^{s} \omega_{c}^{1-s} e^{-\omega / \omega_{c}}$, where $\eta_j$ is the dimensionless coupling strength of the $j$th qubit, $\omega_c$ is the cutoff frequency, and $s$ is the Ohmicity parameter. Depending on the value of $s$, the environment is classified as sub-Ohmic for $0 < s < 1$, Ohmic for $s = 1$, or super-Ohmic for $s > 1$ \cite{RevModPhys.59.1}. Under the initial condition that the three independent environments are in vacuum states, we can exactly trace out the environmental degrees of freedom from the unitary dynamics governed by the total Hamiltonian and obtain a non-Markovian master equation for the reduced density matrix of the three-qubit system
\begin{eqnarray}
&&\dot{\rho}_{ABC}(t) = \sum_{j=A,B,C}\{-i\Omega_j(t)[\hat{\sigma}_{j}^{\dagger}\hat{\sigma}_{j},\rho_{ABC}(t)]+\Gamma_j(t)~~~~\nonumber\\
&&\times[2\hat{\sigma}_{j}\rho_{ABC}(t)\hat{\sigma}_{j}^{\dagger}-\hat{\sigma}_{j}^{\dagger}\hat{\sigma}_{j}\rho_{ABC}(t)-\rho_{ABC}(t)\hat{\sigma}_{j}^{\dagger}\hat{\sigma}_{j}]\},\label{m5}
\end{eqnarray}
where $\Omega_j(t)=-\operatorname{Im}[\dot{u}_j(t) / u_j(t)]$ is the Lamb-shifted frequency and $\Gamma_j(t)=-\operatorname{Re}[\dot{u}_j(t) / u_j(t)]$ is the decay rate. The time-dependent coefficient $u_j(t)$ is determined by 
\begin{equation}
\dot{u}_j(t) + i\omega_{0}u_j(t) + \int_{0}^{t} f_j(t-\tau) u_j(\tau)  d\tau = 0, \label{e4}
\end{equation}
where $f_j(t-\tau)=\int_{0}^{\infty} J_j(\omega)e^{-i\omega(t-\tau)}d\omega$ is the correlation function of $j$th environment. Equation \eqref{e4} indicates that all non-Markovian effects are self-consistently incorporated into the time-dependent coefficients $\Omega_j(t)$ and $\Gamma_j(t)$.

Now consider the initial three-qubit state as a mixed entangled state $\rho_{ABC}(0) = \frac{1 - q}{2}\left( |\chi_1\rangle\langle\chi_1| + |\chi_2\rangle\langle\chi_2| \right) + \frac{ q }{8}\mathbb{I}_8$, where $|\chi_1\rangle = \frac{1}{\sqrt{6}}( |010\rangle - 2|100\rangle + |001\rangle )$,
$|\chi_2\rangle = \frac{1}{\sqrt{6}}( |101\rangle - 2|011\rangle + |110\rangle )$, and
$q \in (0, 1)$ \cite{PhysRevA.94.042105}. According to Eq. \eqref{m5}, we obtain the reduced density matrices $\rho_{AB}(t)$, $\rho_{AC}(t)$, and $\rho_{BC}(t)$ in the energy eigenbasis $\{|gg\rangle, |ge\rangle, |eg\rangle, |ee\rangle\}$ (see Appendix \ref{Appen} for the explicit forms). Therefore, if Alice performs a measurement on her qubit, the centers of Bob's and Charlie's QSEs are given by
\begin{equation}
    \mathbf{c}_{B|A}=\mathbf{c}_{C|A} = (0,\;0,\;1 + \frac{|u_b|^2\bigl[6 - 2p + (2p - 3)|u_a|^2\bigr]}{3(|u_a|^2 - 2)} ),
\end{equation}
and the corresponding semiaxis lengths are
\begin{equation}
\begin{aligned}
r_{B|A}^{x,y} = r_{C|A}^{x,y} &= \frac{2p|u_b|}{3\sqrt{2 - |u_a|^2}}, \\
r_{B|A}^{z} = r_{C|A}^{z} &= \frac{2p|u_b|^2}{3||u_a|^2 - 2 |}.
\label{e8}
\end{aligned}
\end{equation}
Consider now the case where Bob performs the measurement. The centers of Alice's and Charlie's QSEs are
\begin{equation}
\begin{aligned}
\mathbf{c}_{A|B} &= (0, \; 0, \; 1 + \frac{|u_a|^2}{3}  [2p (1 + \frac{1}{|u_b|^2 - 2} )-3 ] ) \\
\mathbf{c}_{C|B} &= (0,\;0,\;
\frac{ (6+p)|u_c|^2 + |u_b|^2[3 - (3+p)|u_c|^2]-6}{3(|u_b|^2 - 2)}
),
\end{aligned}
\end{equation}
and the corresponding semiaxis lengths are
\begin{equation}
\begin{aligned}
r_{A|B}^{x,y}  &= \frac{2p|u_a|}{3\sqrt{2 - |u_b|^2}},~~~r_{A|B}^{z} = \frac{2p|u_a|^2}{3||u_b|^2 - 2|}, \\
r_{C|B}^{x,y}  &= \frac{p|u_c|}{3\sqrt{2 - |u_b|^2}},~~~r_{C|B}^{z} = \frac{p|u_c|^2}{3||u_b|^2 - 2|}.
\end{aligned}
\label{e9}
\end{equation}
Finally, when Charlie is the measuring party, the centers of Alice's and Bob's QSEs are
\begin{equation}
\begin{aligned}
\mathbf{c}_{A|C} &=  (0, \; 0, \; 1 + \frac{|u_a|^2}{3}  [2p (1 + \frac{1}{|u_b|^2 - 2} )-3 ] ), \\
\mathbf{c}_{B|C} &= (0,\;0,\;
\frac{ 3|u_c|^2 + |u_b|^2[6 + p - (3+p)|u_c|^2]-6}{3(|u_c|^2 - 2)}
),
\end{aligned}
\end{equation}
and the corresponding semiaxis lengths are
\begin{equation}
\begin{aligned}
r_{A|C}^{x,y}  &= \frac{2p|u_a|}{3\sqrt{2 - |u_c|^2}},~~~r_{A|C}^{z} = \frac{2p|u_a|^2}{3||u_c|^2 - 2|}, \\
r_{B|C}^{x,y}  &=  \frac{p|u_b|}{3\sqrt{2 - |u_c|^2}},~~~r_{B|C}^{z} = \frac{p|u_b|^2}{3||u_c|^2 - 2|}.
\end{aligned}
\label{e10}
\end{equation}
With these expressions, we can now analyze how the volume monogamy relation behaves under dissipative dynamics. 

In the weak-coupling limit, where the environment's correlation time is much shorter than the characteristic time scale of each qubit, the Born-Markov approximation becomes applicable. Under this approximation, Eq. \eqref{e4} yields an exponentially decaying solution $u_j^{\mathrm{MA}}(t) = \exp[-\kappa_j - i(\omega_0 + \Delta_j)] t$, where $\kappa_j = \pi J_j(\omega_0)$ is the decay rate and $\Delta_j = \mathcal{P} \int_0^\infty \frac{J_j(\omega)}{\omega_0 - \omega} d\omega$ represents the Lamb shift, with $\mathcal{P}$ denoting the Cauchy principal value. When this solution is inserted into Eqs. \eqref{e8}, \eqref{e9}, and \eqref{e10}, all semiaxis lengths $r_{X|Y}^{x,y,z}$ of the QSEs are found to diminish to zero over time, indicating the complete loss of quantum correlations and that the volume monogamy relations reduce to a trivial form. The loss of bipartite correlations resulting from the Markov approximation is similar to the decoherence behavior reported in Refs. \cite{PhysRevA.102.020402,PhysRevA.87.032317,radhakrishnan_time_2019}. 

We now go beyond the Born-Markov approximation. In the general non-Markovian regime, solving Eq. \eqref{e4} requires numerical methods. Nevertheless, the long-time asymptotic behavior of $u_j(t)$ can be obtained analytically via the Laplace transform method. Applying the transform to Eq. \eqref{e4} yields $\tilde{u}_j(z )=[z+i\omega_{0}+\int_{0}^{\infty}\frac{J_j(\omega)}{z+i\omega}d\omega]^{-1}$. Then $u_j(t)$ can be obtained by performing the inverse Laplace transform on $\tilde{u}_j(z)$. This requires finding the poles of $\tilde{u}_j(z)$ from the equation
\begin{equation}
Y_j(E) \equiv \omega_{0} - \int_{0}^{\infty} \frac{J_j(\omega)}{\omega - E} d\omega = E, \quad (E = iz).\label{S5}
\end{equation}
A deeper insight is that the roots of Eq. \eqref{S5} directly correspond to the eigenenergies of each local qubit-environment system in the single-excitation subspace. To see this, we write the eigenstate as $|\Phi_j\rangle = ( c_j \hat{\sigma}_j^\dagger + \sum_k d_{jk} \hat{a}_{jk}^\dagger) |g_j, \{0_{jk}\} \rangle$. Substituting this state into $\hat{H}_j |\Phi_j\rangle = E_j |\Phi_j\rangle$, we obtain the eigenenergy equation $E_j - \omega_0 - \sum_k |g_{jk}|^2/(E_j-\omega_k ) = 0$, which matches Eq. \eqref{S5} in the continuum limit of $k$. The function $Y_j(E)$ is monotonically decreasing for $E<0$; thus, a single isolated root $E_j^b$ exists in this region provided $Y_j(0)<0$. For $E>0$, $Y_j(E)$ is non-analytic and Eq. \eqref{S5} yields infinitely many roots that form a continuous energy band. The eigenstate associated with the isolated root $E_j^b$ is referred to as a bound state. Applying the residue theorem and the inverse Laplace transform, we obtain \cite{PhysRevA.103.L010601}
\begin{equation}
u_j(t)=Z_j e^{-i E_j^{b} t}+\int_{0}^{\infty} \mathcal{C}(E)e^{-i E t} d E.
\end{equation}
Here $Z_j \equiv \big[ 1 + \int_{0}^{\infty} \frac{J_j(\omega)  d\omega}{(E_j^{b} - \omega)^{2}} \big]^{-1}$ is the contribution of the bound state, and $ \mathcal{C}(E)=J_j(E)/\{[E-\omega_{0}-\Delta_j(E)]^{2}+[\pi J_j(E)]^{2}\}$ stems from the continuous energy band. The second term of $u_j(t)$ decays to zero at long times due to out-of-phase interference. Thus, we have
\begin{equation}
    \lim_{t\to\infty} u_j(t)=
    \begin{cases} 
    0; & \text{without bound state} \\ 
    Z_je^{-iE_j^{b}t}; & \text{with bound state}
    \end{cases}.\label{e11}
\end{equation}
The condition for bound-state formation in the Ohmic-family spectral density follows from $Y_j(0) < 0$, yielding $\eta_j > \eta_{c} \equiv \omega_{0} / [\omega_{c} \gamma(s)]$, where $\gamma(s)$ denotes the Gamma function.

Equations \eqref{e8}, \eqref{e9}, and \eqref{e10}, together with Eq. \eqref{e11}, clearly indicate the crucial role played by the bound state. In the absence of bound states, $u_j(t)$ decays to zero, causing the corresponding QSE to collapse to a single point within the Bloch sphere, as in the Born-Markov case \cite{PhysRevA.91.022301,PhysRevA.102.020402}. Conversely, the emergence of a bound state on a given side protects the semiaxis lengths of that party's QSE from decay and sustains a finite structure. Hence, the QSE features of Alice, Bob, and Charlie, and thus the volume monogamy relation, are determined by whether each side supports a bound state. This observation reveals how volume monogamy behaves in open systems. Bound states sustain nontrivial monogamy under decoherence, while their absence makes it trivial. This inspires a new perspective. By selectively engineering bound states, we can realize extreme volume monogamy, i.e., $v_{C|A}=0$, while preserving steering between trusted parties. In the following, we focus on the steering ellipsoids generated by trusted parties (Alice and Bob) as the measurers, since Charlie is untrusted and thus is not considered as a secure measurer. By designing which parties support bound states, we can selectively preserve steering between trusted parties while completely eliminating the steerability of the untrusted party. Below, we examine each of these cases in detail.

\section{numerical results}\label{3}
\begin{figure}
\hspace*{-6pt}\includegraphics[width=1.03\columnwidth]{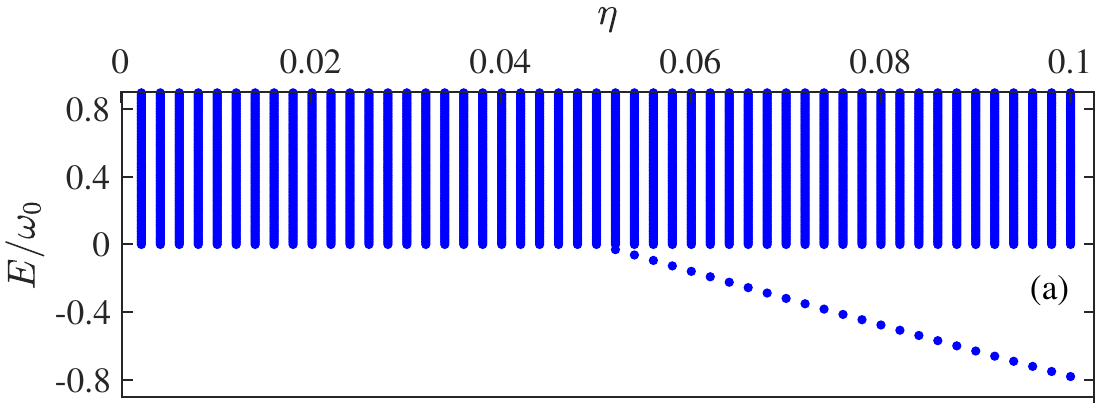}
\includegraphics[width=\columnwidth]{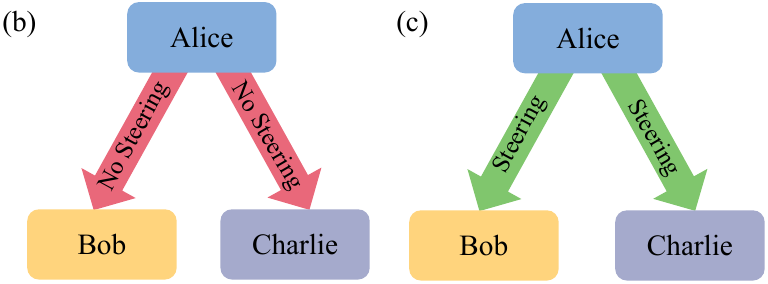}
\hspace*{5pt}\includegraphics[width=\columnwidth]{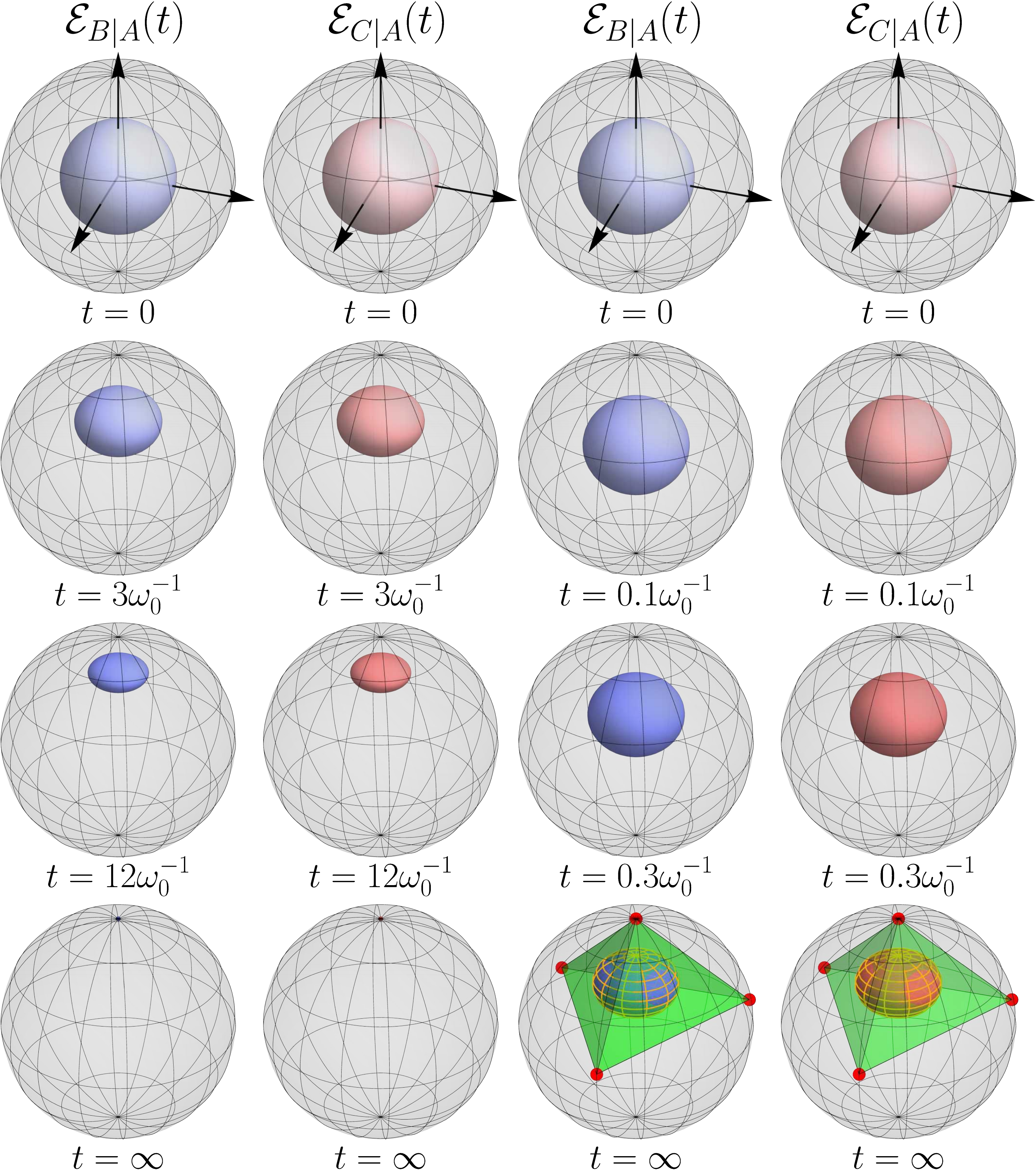}
\caption{(a) Energy spectrum of each qubit-environment subsystem. (b) Steering diagram with Alice as the measurer, showing the steering from Alice to Bob and Charlie, together with the evolution of $\mathcal{E}_{B|A}$ and $\mathcal{E}_{C|A}$ from top to bottom for no bound states $\eta_A = \eta_B = \eta_C = 0.03$. (c) Same as (b), but for bound states on all sides with $\eta_A = \eta_B = \eta_C = 0.06$. The solid surfaces show the numerical evolution, while the golden wireframes on the final QSEs represent the analytical steady-state solutions. The green tetrahedron represents any tetrahedron inscribed in the Bloch sphere. In all plots, we use $\omega_c=20\omega_0$ and $q=3/4$.} \label{FIG1}
\end{figure}

\begin{figure}
\includegraphics[width=\columnwidth]{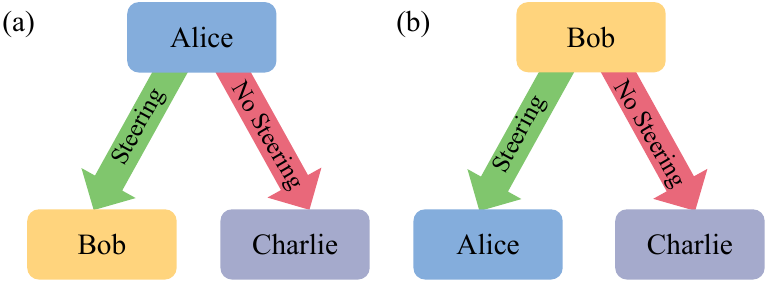}
\hspace*{6pt}\includegraphics[width=\columnwidth]{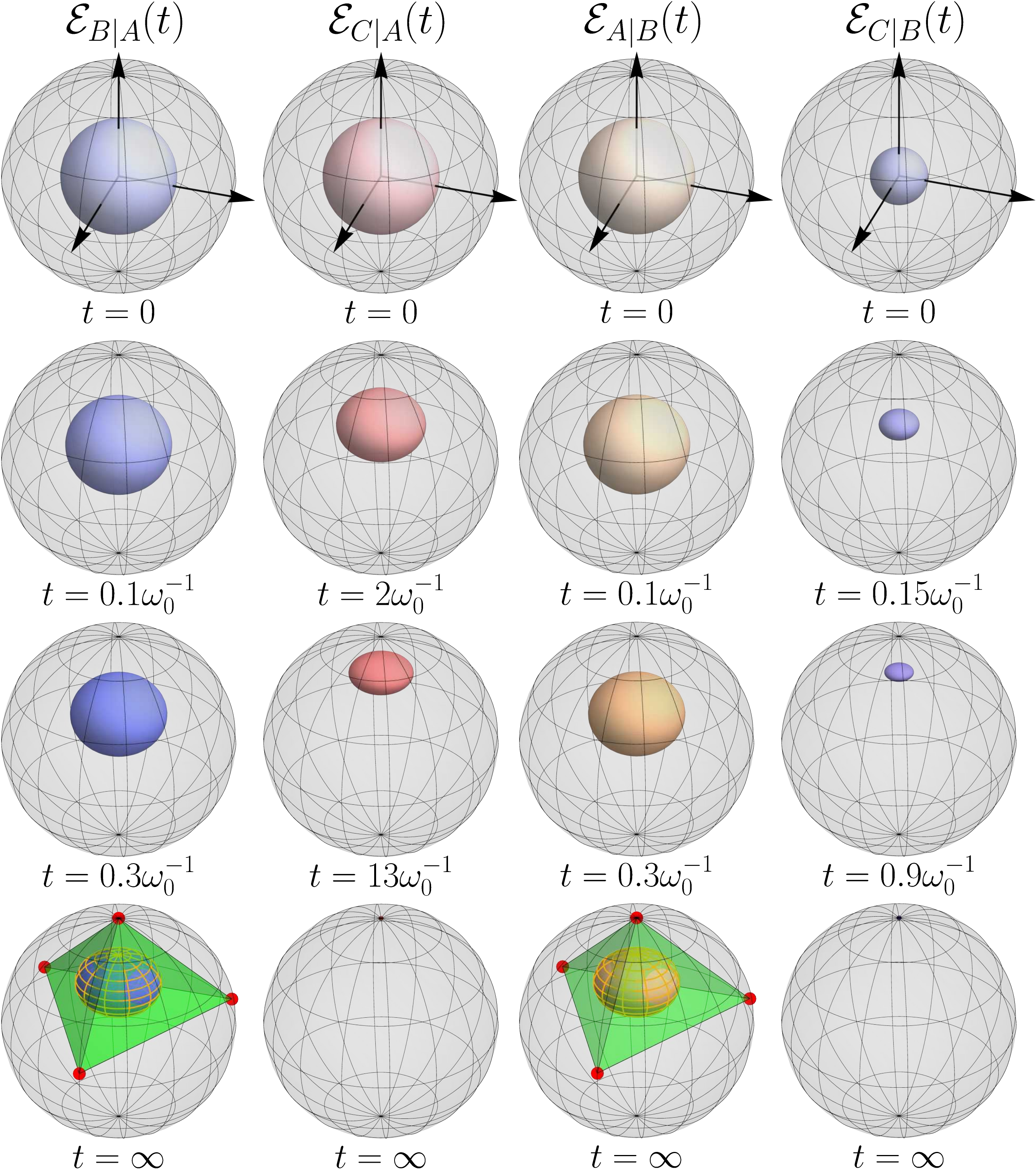}
\caption{(a) Steering diagram with Alice as the measurer, showing the steering from Alice to Bob and Charlie, together with the evolution of $\mathcal{E}_{B|A}$ and $\mathcal{E}_{C|A}$ from top to bottom for bound states on Alice and Bob but absent on Charlie, i.e., $\eta_A = \eta_B = 0.06$ and $\eta_C = 0.03$. (b) Same as (a), but with Bob as the measurer, showing the steering from Bob to Alice and Charlie. The solid surfaces are numerical results, and the golden wireframes are analytical solutions. The green tetrahedron cannot enclose the steady QSE, indicating entanglement. In all plots, we use $\omega_c=20\omega_0$ and $q=3/4$.} \label{FIG2}
\end{figure}
To verify our analytical results, we choose an Ohmic spectral density ($s = 1$) and set $q = 3/4$ in the initial state, which supports tripartite entanglement. Figure \ref{FIG1}(a) depicts the energy spectrum of each qubit and its local environment, where a bound state appears once $\eta > \eta_c = 0.05$. We first examine the case where no bound state is present on any side, taking $\eta_A = \eta_B = \eta_C = 0.03$. Figure \ref{FIG1}(b) illustrates the steering scenario when Alice acts as the measurer. The solid blue and red surfaces show the evolution of the QSEs $\mathcal{E}_{B|A}$ and $\mathcal{E}_{C|A}$, respectively, obtained from the exact numerical solution of Eq.~\eqref{e4}. Both QSEs shrink to a single point in the long-time limit due to decoherence, indicating that the volume monogamy relation becomes trivial. This complete collapse confirms our analytical predictions and is consistent with the Markovian result. Similarly, when Bob performs the measurement, the same collapse occurs for the QSEs, although the results are not shown here. In contrast, when all three parties support bound states, we set $\eta_A = \eta_B = \eta_C = 0.06$. Figure \ref{FIG1}(c) also shows the case where Alice is the measurer. The QSEs $\mathcal{E}_{B|A}$ and $\mathcal{E}_{C|A}$ evolve into steady structures in the long-time limit, in sharp contrast to the complete collapse observed in the case without bound states. These steady states are in excellent agreement with the analytical predictions obtained from Eqs.~\eqref{e8}, \eqref{e9}, \eqref{e10}, and \eqref{e11}, as illustrated by the golden wireframes overlaid on the final QSEs. This confirms the essential role of bound states in preserving the QSEs and ensuring that the volume monogamy relation remains nontrivial under decoherence. Moreover, the steady-state QSE cannot be enclosed by any tetrahedron inscribed in the Bloch sphere, indicating bipartite entanglement in the $AB$ and $AC$ subsystems according to the nested tetrahedron condition \cite{PhysRevLett.113.020402}. Although this configuration does not eliminate Charlie's information access, it protects the entanglement between Alice and Bob and maintains the conventional volume monogamy relation in a noisy environment. Next, we aim to completely eliminate Charlie's information access.

We now consider the configuration where bound states are present on the trusted parties (Alice and Bob) but absent on Charlie, i.e., $\eta_A, \eta_B=0.06$ while $\eta_C =0.03$. When Alice acts as the measurer, the numerical evolution of $\mathcal{E}_{B|A}$ and $\mathcal{E}_{C|A}$ is shown in Fig. \ref{FIG2}(a). The QSE $\mathcal{E}_{B|A}$ maintains a finite volume in the steady state, whereas $\mathcal{E}_{C|A}$ completely collapses to a point in the long-time limit. Since the QSE volume quantifies the steerability of a party and thus reflects how much information it can access, selective bound-state engineering allows Bob to access information from Alice’s measurements while completely blocking Charlie. The crucial role of bound states is thus further confirmed. Similarly, when Bob performs the measurement, $\mathcal{E}_{A|B}$ remains finite while $\mathcal{E}_{C|B}$ collapses to a point ($v_{C|B}=0$), as shown in Fig.~\ref{FIG2}(b). This means that Bob, as the measurer, can also prevent information leakage to Charlie and steer to Alice. Moreover, the steady-state $\mathcal{E}_{B|A}$ and $\mathcal{E}_{A|B}$ cannot be enclosed by any tetrahedron inscribed in the Bloch sphere, indicating the presence of bipartite entanglement in the $AB$ subsystem according to the nested tetrahedron condition \cite{PhysRevLett.113.020402}. The above analysis shows that the two-sided bound-state protection mechanism preserves two-way quantum steering with entanglement, which lays the foundation for two-way quantum communication. The golden wireframes overlaid obtained from Eqs.~\eqref{e8}, \eqref{e9}, \eqref{e10}, and \eqref{e11} on the final QSEs confirm excellent agreement between the analytical predictions and numerical simulations. This configuration thus completely eliminates the untrusted party's information accessibility while preserving the bipartite entanglement between Alice and Bob, which is essential for two-way quantum communication in tripartite quantum networks.
\begin{figure}
\includegraphics[width=\columnwidth]{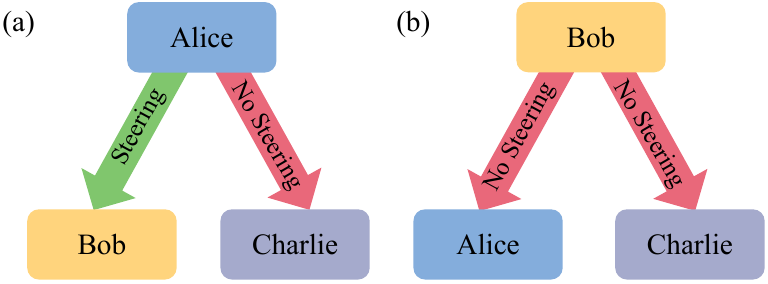}
\hspace*{6pt}\includegraphics[width=\columnwidth]{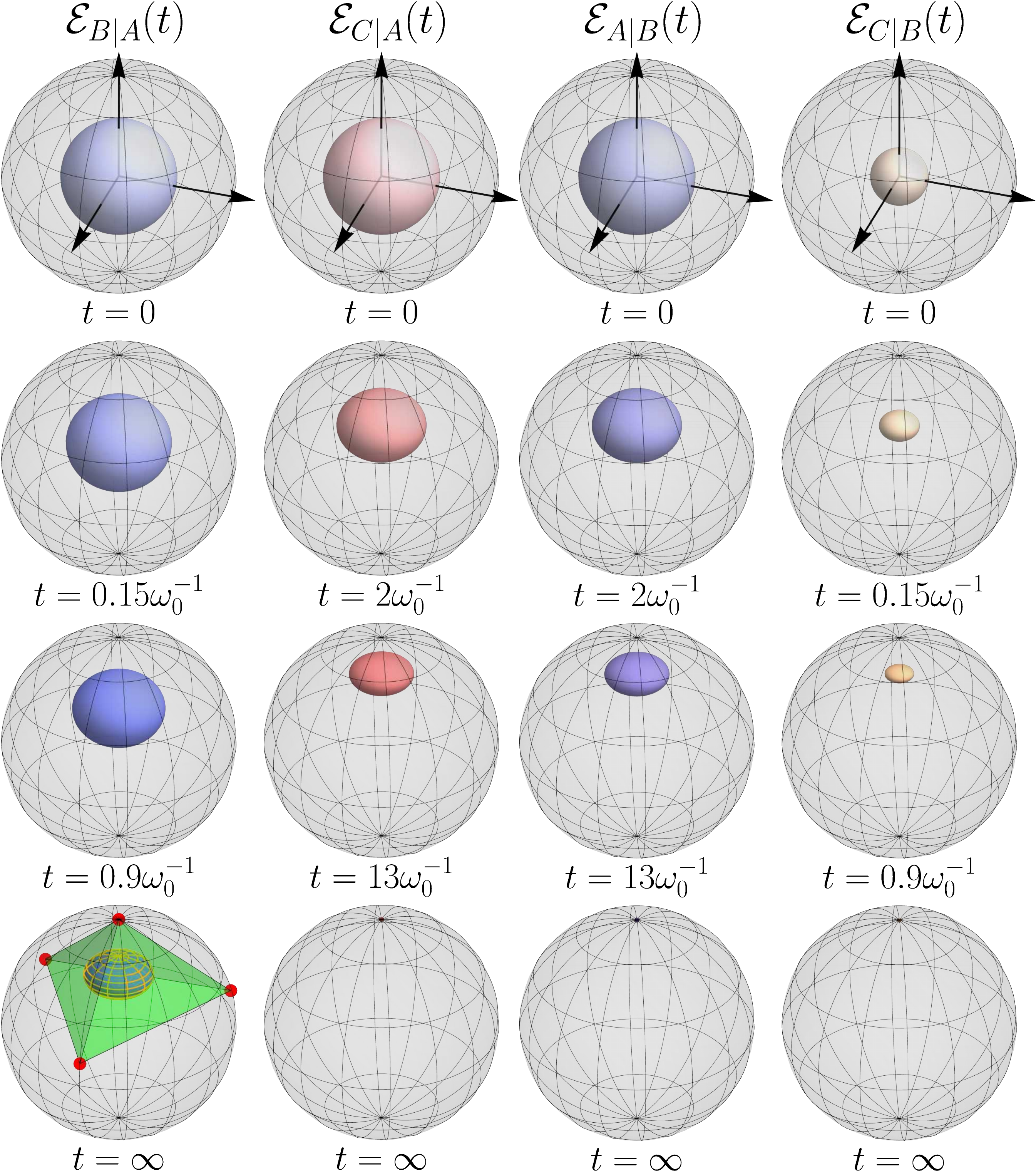}
\caption{(a) Steering diagram with Alice as the measurer, showing the steering from Alice to Bob and Charlie, together with the evolution of $\mathcal{E}_{B|A}$ and $\mathcal{E}_{C|A}$ from top to bottom for a bound state only on Bob, i.e., $\eta_B = 0.06$ and $\eta_A = \eta_C = 0.03$. (b) Same as (a), but with Bob as the measurer, showing the steering from Bob to Alice and Charlie. The solid surfaces are numerical results, and the golden wireframes are analytical solutions. The green tetrahedron can enclose the steady QSE, indicating separability. In all plots, we use $\omega_c=20\omega_0$ and $q=3/4$.} \label{FIG3}
\end{figure}

For the case where only one trusted party supports a bound state. Without loss of generality, we first assume that Bob supports a bound state while Alice and Charlie do not, i.e., $\eta_B = 0.06$ and $\eta_A = \eta_C = 0.03$. When Alice performs the measurement, the numerical evolution of $\mathcal{E}_{B|A}$ and $\mathcal{E}_{C|A}$ is shown in Fig.~\ref{FIG3}(a). $\mathcal{E}_{B|A}$ is protected by his bound state and thus maintains a finite volume in the steady state, while $\mathcal{E}_{C|A}$ collapses to a single point. This indicates that Alice can still steer Bob, but cannot steer Charlie. Since steerability directly determines how much information a party can extract, this configuration allows Bob to obtain information from Alice's measurements while Charlie is completely blocked. When Bob acts as the measurer, as shown in Fig. \ref{FIG3}(b), both $\mathcal{E}_{A|B}$ and $\mathcal{E}_{C|B}$ collapse completely, meaning that in this configuration Bob can steer neither Alice nor Charlie. Further analysis shows that the surviving $\mathcal{E}_{B|A}$ lies entirely inside a tetrahedron inscribed in the Bloch sphere, indicating that $\rho _{AB}$ is a separable state according to the nested tetrahedron condition \cite{PhysRevLett.113.020402}. This demonstrates that a single bound state on Bob enables only one-way quantum steering based on a separable state: Alice can steer Bob, but Bob cannot steer Alice. This is because Alice lacks a bound state, and a single bound state on Bob alone is insufficient to sustain entanglement with Alice. The case where only Alice supports a bound state is analogous and is not shown here. When Bob acts as the measurer, $\mathcal{E}_{A|B}$ remains finite while $\mathcal{E}_{C|B}=0$, enabling one-way steering from Bob to Alice; when Alice acts as the measurer, both $\mathcal{E}_{B|A}$ and $\mathcal{E}_{C|A}$ collapse to points. Thus, a single bound state can only protect the QSE on that side, enabling one-way quantum steering only from the other party to the protected one. The numerical steady-state QSEs agree well with the analytical predictions, as shown by the golden wireframes in the figures. Thus, the one-sided bound state configuration eliminates the untrusted party's information accessibility, at the cost of allowing only one-way quantum steering.
\section{Discussion and conclusion}\label{4}
It is worth noting that our bound-state based protection mechanism does not rely on the specific form of the spectral density. Although we have employed the Ohmic-family spectral density for concreteness, the underlying physics applies to other spectral structures as well. The key parameters of the spectral density $J(\omega)$ can be experimentally controlled via quantum reservoir engineering techniques \cite{PhysRevA.109.062402,andersen_strongly_2011,PhysRevB.95.161408,PhysRevA.84.031602}. Moreover, bound states and their associated dynamical effects have already been observed in state-of-the-art quantum optics platforms, including circuit quantum electrodynamics \cite{liu_quantum_2017} and matter-wave systems \cite{krinner_spontaneous_2018}. These experimental developments provide a strong support for our theoretical investigations. 

In summary, we have investigated the protection and control of QSE in open multipartite systems through selective bound-state engineering. We found that the geometry of each party's QSE is jointly determined by non-Markovian effects and the formation of bound states. The presence of a bound state on a given side protects the corresponding QSE from decoherence, while its absence leads to complete collapse. Based on this mechanism, we proposed a scheme to realize extreme volume monogamy of quantum steering in open systems, defined as the condition $v_{C|A}=0$, thereby eliminating the untrusted party's information accessibility while preserving steering between trusted parties. By engineering which parties support bound states in their local environments, we systematically examined four distinct configurations: three-sided bound states, two-sided bound states, one-sided bound states, and no bound states. When no bound states are present on any side, all QSEs collapse to points under decoherence, consistent with the Markovian result. In contrast, when all three parties support bound states, the QSEs are protected and the volume monogamy relation remains nontrivial, although Charlie's information accessibility is not eliminated. When bound states are present on both trusted parties (Alice and Bob) but absent on the untrusted party (Charlie), we achieve extreme volume monogamy. This configuration thus completely eliminates the untrusted party's information leakage while protecting the entanglement essential for two-way quantum communication between the trusted parties. When only one trusted party supports a bound state, the scheme still eliminates Charlie's information accessibility, but only enables one-way quantum steering. This configuration may find applications in asymmetric quantum communication scenarios where only one-way steering is required. Our results establish selective bound-state engineering as an effective mechanism for realizing extreme volume monogamy of quantum steering in open systems. The ability to selectively preserve steering between trusted parties while completely eliminating it for an untrusted party provides a basis for secure quantum communication with an untrusted third party in noisy environments.

\begin{acknowledgments}
The work is supported by the National Natural Science Foundation of China (Grants No. 12275109, No. 92576202, and No. 12247101), the Quantum Science and Technology-National Science and Technology Major Project (Grant No. 2023ZD0300904), the Gansu Science and Technology Leading Talent Program (Grant No. 26RCKA011 ), the Natural Science Foundation of Gansu Province (Grant No. 25JRRA799), and the Fundamental Research Funds for the Central Universities (Grant No. lzujbky-2025-jdzx07). 
\end{acknowledgments}

\begin{appendix}
\section{Explicit expressions for $\rho_{AB}(t)$, $\rho_{AC}(t)$, and $\rho_{BC}(t)$}\label{Appen}

According to Eq.~\eqref{m5}, the reduced density matrix $\rho_{AB}(t) = \sum_{m,n;m',n'} \rho_{AB}^{mm',nn'}(t)|m,n\rangle\langle m',n'|$, where
\begin{equation}
\begin{aligned}
\rho_{AB}^{gg,gg}(t) &= \frac{1}{12}[-6(-2+|u_b(t)|^2) \\
&\quad + |u_a(t)|^2(-6+(3-2p)|u_b(t)|^2)],\\[4pt]
\rho_{AB}^{ge,ge}(t) &= \frac{1}{12}[6+(-3+2p)|u_a(t)|^2]|u_b(t)|^2,\\
\rho_{AB}^{eg,eg}(t) &= \frac{1}{12}|u_a(t)|^2[6+(-3+2p)|u_b(t)|^2],\\
\rho_{AB}^{ee,ee}(t) &= \frac{1}{12}(3-2p)|u_a(t)|^2|u_b(t)|^2,\\
\rho_{AB}^{ge,eg}(t) &= -\frac{p}{3}u_b(t)u_a^*(t),
\end{aligned}
\end{equation}
and all other elements are zero.

For $\rho_{AC}(t) = \sum_{m,n;m',n'} \rho_{AC}^{mm',nn'}(t)|m,n\rangle\langle m',n'|$, the nonzero matrix elements are
\begin{equation}
\begin{aligned}
\rho_{AC}^{gg,gg}(t) &= \frac{1}{12}[-6(-2+|u_c(t)|^2) \\
&\quad + |u_a(t)|^2(-6+(3-2p)|u_c(t)|^2)],\\[4pt]
\rho_{AC}^{ge,ge}(t) &= \frac{1}{12}[6+(-3+2p)|u_a(t)|^2]|u_c(t)|^2,\\
\rho_{AC}^{eg,eg}(t) &= \frac{1}{12}|u_a(t)|^2[6+(-3+2p)|u_c(t)|^2],\\
\rho_{AC}^{ee,ee}(t) &= \frac{1}{12}(3-2p)|u_a(t)|^2|u_c(t)|^2,\\
\rho_{AC}^{ge,eg}(t) &= -\frac{p}{3}u_c(t)u_a^*(t),
\end{aligned}
\end{equation}
and all other elements are zero.

For $\rho_{BC}(t) = \sum_{m,n;m',n'} \rho_{BC}^{mm',nn'}(t)|m,n\rangle\langle m',n'|$, the nonzero matrix elements are
\begin{equation}
\begin{aligned}
\rho_{BC}^{gg,gg}(t) &= \frac{1}{12}[-6(-2+|u_c(t)|^2) \\
&\quad + |u_b(t)|^2(-6+(3+p)|u_c(t)|^2)],\\[4pt]
\rho_{BC}^{ge,ge}(t) &= -\frac{1}{12}[-6+(3+p)|u_b(t)|^2]|u_c(t)|^2,\\
\rho_{BC}^{eg,eg}(t) &= -\frac{1}{12}|u_b(t)|^2[-6+(3+p)|u_c(t)|^2],\\
\rho_{BC}^{ee,ee}(t) &= \frac{1}{12}(3+p)|u_b(t)|^2|u_c(t)|^2,\\
\rho_{BC}^{ge,eg}(t) &= \frac{p}{6}u_c(t)u_b^*(t),
\end{aligned}
\end{equation}
and all other elements are zero.
\end{appendix}

\bibliography{ref}
\end{document}